\documentclass[11pt,aps,prc,onecolumn,nofootinbib,superscriptaddress]
{revtex4-2}
\usepackage{amssymb,mathrsfs,xcolor}
\usepackage{amsmath,amsfonts,graphicx}
\usepackage{bm} 

\def\bi{\bibitem}
\usepackage{amssymb,mathrsfs,xcolor}

\newcommand{\bsig}{\mbox{\boldmath $\sigma$}}

\newcommand{\pp}{{\bf p}}
\newcommand{\vv}{{\bf v}}

\newcommand{\la}{\langle}
\newcommand{\ra}{\rangle}

\newcommand{\be}{\begin{equation}}
\newcommand{\ee}{\end{equation}}
\newcommand{\nn}{\nonumber\\}
\newcommand{\eqn}[1]{\label{#1}}
\newcommand{\eq}[1]{Eq.~(\ref{#1})}

\newcommand{\lb}{\label}
\newcommand{\bea}{\begin{eqnarray}}
\newcommand{\eea}{\end{eqnarray}}

\begin{document}

\title{Perturbation theory for a systematic account \\
of the bound-state motion}

\author{Alexander~N.~Kvinikhidze}
\email{sasha\_kvinikhidze@hotmail.com}
\affiliation{Andrea Razmadze Mathematical Institute of Tbilisi
State University,\\
6, Tamarashvili Str., 0186 Tbilisi, Georgia}
\affiliation{College of Science and Engineering, Flinders University,
Bedford Park, SA 5042, Australia} 

\author{Hagop Sazdjian}
\email{hagop.sazdjian@ijclab.in2p3.fr}
\affiliation{Université Paris-Saclay, CNRS/IN2P3, IJCLab,
91405 Orsay, France}  

\author{Boris Blankleider}
\email{boris.blankleider@flinders.edu.au}
\affiliation{College of Science and Engineering, Flinders University,
Bedford Park, SA 5042, Australia} 



\begin{abstract}
We derive a perturbation theory (PT) for the Lorentz boost operator
in the space of two-nucleon wave functions. The latter is expressed
in terms of the nucleon-nucleon ($NN$) potentials, developed so far
in great detail for their use in the $NN$ scattering studies.
The PT is designed to take into account the boost relativistic
corrections in a systematic way and, as such, it is the only missing
part in the corresponding approaches developed up to now in the
low-energy effective field theories.
\end{abstract}

\maketitle

\section{Introduction} \lb{s1} 

In theoretical studies of the few-body bound state problems (in general,
with any amplitude of processes involving moving bound states) it
is important to express them in terms of the bound state wave functions
at rest, because the latter are more convenient for calculations
than those of the moving bound states. 
These include the study of many characteristic features of the bound
states' structure, such as their size, shape, etc.
\par
In covariant approaches,  
the wave functions in different reference frames
are related to each other simply by Lorentz transformations of
space-time or energy-momentum variables. Unfortunately, in the most
convenient approaches to the extensively used Weinberg's effective
field theory (EFT) \cite{Wein,vKolck92,vKolck94},
time-ordered perturbation theory (TOPT)
and the so-called method of unitary transformation (MUT) (where 
a unitary transformation of the pion-nucleon Hamiltonian is used to
decouple the purely nucleonic subspace of the Fock space from the rest),
the Lorentz boost transformations are dynamical, i.e., they depend on
the interaction. 
This is not  a problem when nonrelativistic Galilei boosting of wave
functions suffices, but the currently pursued high precision
calculations require from the relativistic boost
corrections (including the above mentioned interaction part) to be
accounted for. This is the task addressed in the present paper. 
\par    
Our approach is based on the development of a perturbation theory
(PT) taking a systematic account of the bound state motion in the
EFT \cite{UT-boost,koling}.
Generally, in the literature, the account of the bound state motion 
is done in a non-systematic way, based on approximate formulas,
without having prescriptions for higher-order corrections.
PT is mostly useful for practitioners of the extensively used
approaches to chiral effective field theories, including those based
on the TOPT and the MUT methods.
\par
In this view it is worth mentioning that a PT for the quantized
boost operator has been derived in the 1980s
\cite{KK-89,KMK-88-boost,KMagK-89}, whose diagram technique
generates functions that, in contrast to the ones presented below,
are not directly related to the $NN$ potentials.\footnote{This PT
has been derived for the expansion of the quantized boost operator
with respect to its interaction part, which at once is the expansion
in the boosting speed, just as the well known evolution operator
$E^{it(H_0+H_I)}$ is expanded in the interaction $H_I$ and the
evolution time $t$ simultaneously. However, the corresponding
diagram technique generates functions which are determined by the
vertices given in the Lagrangian, rather than directly by the $NN$
potentials.} 
\par
The plan of the paper is the following. In Sec. \ref{New}, we
develop the PT starting from the relativistic
Schr\"{o}dinger equation. In Sec. \ref{s3},
we display the kinematic part of the Lorentz
boost operator for the case of moving bound states. In Sec. \ref{s4},
a generalized form of the PT is presented, in which the unperturbed
equation includes a part of the Lorentz boost transformation.
Conclusion is presented in Sec. \ref{s5}. Three Appendixes present
technical details of related calculations.
\par

\section{PT based on the relativistic Schr\"{o}dinger equation}
\label{New}
 
The construction of the PT is based on the relativistic Schr\"{o}dinger
equation for the moving bound state,
\be
\left(P^0-\sqrt{m_1^2+{\bf p}_1^2}-\sqrt{m_2^2+{\bf p}_2^2}\right)
\Psi_P({\bf p})=\int \frac{d{\bf k}}{(2\pi)^3}V(P,{\bf p},{\bf k})
\Psi_P({\bf k}),  \eqn{cor-PT}
\ee
where the (effective) potential $V(P,{\bf p},{\bf k})$ is constructed in
QFT in such a way that in the case of exact $V$ this equation has a
solution at $P^0=\sqrt{M^2+{\bf P}^2}$, where $M$ is the eigenvalue
of \eq{cor-PT} considered in the center-of-mass reference frame,
${\bf P}=0$. The potential $V(P,{\bf p},{\bf k})$ is an effective
potential obtained after eliminating the higher Fock components, so
that it is related to the underlying quantized Hamiltonian,
$H=H_0+H_I$, as follows:
\be
(2\pi)^3 \delta^3({\bf P}-{\bf K})V(P,{\bf p},{\bf k})=
\frac{1}{\sqrt{4\omega({\bf p_1})\omega({\bf p_2})}}
\la p_1,p_2|W(P^0) |k_1,k_2\ra\frac{1}
{\sqrt{4\omega({\bf k_1})\omega({\bf k_2})}},
\eqn{Psi-W}
\ee
where $\la p_1,p_2|W|k_1,k_2\ra$ is defined as the sum of all two-body
irreducible diagrams in a well-known way, by using the two-body Fock
state projection operator ${\cal P}$,
\be
\la p_1,p_2|W(P^0)|k_1,k_2\ra=
\la p_1,p_2|H_I\left(1-\frac{1-{\cal P}}{P^0-H_0}H_I\right)^{-1}
|k_1,k_2\ra.  \eqn{Proj}
\ee
The $\omega$s in \eq{Psi-W} are defined as
$\omega({\bf p}_a^{})=p_a^0=\sqrt{m_a^2+
{\bf p}_a^2}\equiv \omega_a^{}$, $a=1,2$.
\par
The wave function $\Psi_P({\bf p})$ is defined as the projection
of the corresponding eigenstate $|P\ra\ra$ of the quantized
Hamiltonian $H$ on the two-body sector of the Fock space.
To establish the relationship between the wave function
$\Psi_P({\bf p})$ and the corresponding eigenstate $|P\ra\ra$ of
the quantized Hamiltonian $H$, one needs to choose the normalization
condition for $\Psi_P({\bf k})$.
The often used normalization condition
\cite{Lepg,BodwYenn,NewPT,FaustKhel69,Faust70,Sazdj88,JallSazdj97}, 
\be
\int \frac{d{\bf p}}{(2\pi)^3}\frac{d{\bf k}}{(2\pi)^3}
\bar\Psi_P({\bf p}) \left[(2\pi)^3\delta^3({\bf p}-{\bf k})
-\frac{\partial V(P,{\bf p},{\bf k})}{\partial P^0}\right]
\Psi_P({\bf k})=1,  \eqn{Renorm}
\ee
leads to such a relation:
\be \lb{e2*}
(2\pi)^3\delta^3({\bf p}_1+{\bf p}_2-{\bf P})\Psi_P({\bf p})=
\frac{1}{\sqrt{\Omega({\bf p}_1,{\bf p}_2,{\bf P})}}
\la p_1,p_2|P\ra\ra,   
\ee
where we have defined
\be \lb{Omg}
\Omega({\bf p}_1,{\bf p}_2,{\bf P})=
8\omega({\bf p}_1^{})\omega({\bf p}_2^{})\omega_M^{}({\bf P}),
\ \ \ \ \ \ \omega_M^{}({\bf P})=P^0=\sqrt{M^2+{\bf P}^2},  
\ee  
and used the covariant normalization condition of eigenstates of
$H_0$ and $H$ ($|p,a\ra$ and $|P,a\ra\ra$, respectively),
\be
\la k,b|p,a\ra=2\omega({\bf p})(2\pi)^3
\delta^3({\bf p}-{\bf k})\delta_{ab},\hspace{5mm}
\la\la K,b|P,a\ra\ra=2\omega_M({\bf P})(2\pi)^3
\delta^3({\bf P}-{\bf K})\delta_{ab}.
\ee
\par
We remind that the normalization condition (\ref{Renorm}) follows
from the inhomogeneous equation of the Green function $G$,
corresponding to the homogeneous \eq{cor-PT},
\bea \lb{e3}
& &\left(P^0-\sqrt{m_1^2+{\bf p}_1^{\prime 2}}
-\sqrt{m_2^2+{\bf p}_2^{\prime 2}}\right)G(P,{\bf p}',{\bf p})
-\int \frac{d{\bf k}}{(2\pi)^3}
V(P,{\bf p}',{\bf k})G(P,{\bf k},{\bf p})
=(2\pi)^3\delta^3({\bf p}'-{\bf p}),\nonumber \\
& &
\eea
upon use of the bound state pole contribution to $G$, 
\be \lb{e4}
G(P,{\bf p}',{\bf p})=G^b(P,{\bf p}',{\bf p})
+\frac{\,\Psi_P({\bf p}')\bar\Psi_P({\bf p})}{P^0-\sqrt{M^2+{\bf P}^2}}.  
\ee
\par
In view of \eq{e2*}, it may be useful to mention the connection of
$G$ with the quantized Hamiltonian $H$:
\be
(2\pi)^3\delta^3({\bf P}-{\bf P}') G(P,{\bf p},{\bf p}')=
\frac{1}{\sqrt{4 \omega_1 \omega_2}}
\la p_1,p_2|\frac{1}{P^0-H}|p'_1,p'_2\ra
\frac{1}{\sqrt{4 \omega'_1 \omega'_2}},  \eqn{66***}
 \ee
which is clear from  the inhomogeneous part of \eq{e3},
written in the form
\be
G(P,{\bf p},{\bf p}')
=\frac{(2\pi)^3\delta^3({\bf p}-{\bf p}')}{P^0-\omega_1- \omega_2}
+\frac{1}{P^0-\omega_1- \omega_2}
\int \frac{d{\bf k}}{(2\pi)^3} V(P,{\bf p},{\bf k})
G(P,{\bf k},{\bf p}').
\eqn{1112}
\ee  
One can verify that \eq{e2*} is consistent with \eq{66***}.
\par
In actual calculations to some order of the chiral PT, the
dispersion relation $P^0=\sqrt{M^2+{\bf P}^2}$ is violated in the form
$P^0=\sqrt{M^2+{\bf P}^2}+h({\bf P})$;
the function $h({\bf P})$ can be
ignored in what follows, because it is of an order that lies beyond the
considered
approximation.\footnote{By strictly defined rules of construction of
the potential $V$, the corresponding Green function, i.e., the one
which satisfies the equation $G(P)=G_0(P)+G_0(P)V(P)G(P)$, is related
to the quantized Hamiltonian 
$\hat H=\sqrt{\hat{\bf P}^2+M^2}$ as 
$(2\pi)^3\delta^3({\bf P}-{\bf K})G(P)=\la {\bf p}_1,{\bf p}_2|
(P^0-\hat H)^{-1}|{\bf k}_1,{\bf k}_2\ra$.
But the resolvent $\la {\bf p}_1,{\bf p}_2|(P^0-\hat H)^{-1}|
{\bf k}_1,{\bf k}_2\ra$
has a pole just at $P^0=\sqrt{M^2+{\bf P}^2}$, because the quantized
bound state vector $|{P\bf }\ra\ra$ is defined as an eigenvector of
the quantized equation
$\hat P^\mu |{P\bf }\ra\ra=P^\mu |{P\bf }\ra\ra$, which has a solution
only for $P^0=\sqrt{M^2+{\bf P}^2}$, as required by Poincar\'e
invariance.}
The ${\bf P}$ dependence of \eq{cor-PT} makes it difficult to solve it,
that is why a PT representing an expansion of its solution in powers
of the total three-momentum ${\bf P}={\bf p}_1+{\bf p}_2$ is proposed,
in combination with the chiral PT.
Mathematically, the terms of such an expansion depend on the choice of
the second three-dimensional (3D) variable, say ${\bf p}$,
combination of the two momenta ${\bf p}_1$ and ${\bf p}_2$, yielding
$\Psi_{\bf P}({\bf p})=\sum_n C_n({\bf p}){\bf P}^n$. The natural
choices for $\mathbf{p}$ could be the nonrelativistic or the
relativistic relative momentum. We consider here the simplest case,
that of the nonrelativistic relative momentum,
${\bf p}=\frac{m_2{\bf p}_1-m_1{\bf p}_2}{(m_1+m_2)}$.
Different choices of ${\bf p}$ result in different
perturbations $\Delta$ [cf. Eqs. (\ref{Delta}) and (\ref{Delta-n})
below].
\par 
The problem is then reduced to the nonperturbative treatment of this
equation only in the center-of-mass reference frame. Having found
$\Psi_{\bf 0}({\bf p})$ nonperturbatively, $\Psi_{\bf P}({\bf p})$
is looked for as an expansion around  $\Psi_{\bf 0}({\bf p})$,
\be
\Psi_{\bf P}({\bf p})= \Psi_{\bf 0}({\bf p})
+\delta\Psi_{\bf P}({\bf p}),
\eqn{non-relLO}
\ee
where 
$\Psi_{\bf 0}({\bf p})$ satisfies the relativistic Schr\"{o}dinger
equation
\be
\left(M-\sqrt{m_1^2+{\bf p}^2}-\sqrt{m_2^2+{\bf p}^2}\right)
\Psi_{\bf 0}({\bf p})=\int \frac{d{\bf k}}{(2\pi)^3}
V(M,{\bf 0},{\bf p},{\bf k})\Psi_{\bf 0}({\bf k}). 
\eqn{rel-Psi-0}
\ee
The choice of \eq{rel-Psi-0} suits our approach in terms of providing
the unperturbed solution, because such a relativistic equation
is used to study the deuteron in the chiral effective field theory,
at least since 2005 (cf. Eq. (4.1) in \cite{epel5}).
It is then straightforward to derive the PT
for the solution to \eq{cor-PT}, written in the form
\begin{align}
&\left(M-\sqrt{m_1^2+{\bf p}^2}-\sqrt{m_2^2+{\bf p}^2}\right)
\Psi_{\bf P}({\bf p})-
\int \frac{d{\bf k}}{(2\pi)^3}V(M,{\bf 0},{\bf p},{\bf k})
\Psi_{\bf P}({\bf k})\nn
&\hspace{4 cm}=\int \frac{d{\bf k}}{(2\pi)^3}
\Delta(P,{\bf p},{\bf k})\Psi_{\bf P}({\bf k}), 
\eqn{rel-Psi-0-P}
\end{align}
as an expansion around the nonperturbative solution to the
relativistic equation (\ref{rel-Psi-0}) for the bound state wave
function at rest \cite{Lepg,BodwYenn,NewPT},
\be
\Psi_P({\bf p}) = \int\frac{d{\bf k}}{(2\pi)^3}
\la {\bf p}|[1-G^b_u(M) \Delta]^{-1}|
{\bf k}\ra \Psi_{\bf 0}({\bf k}), \eqn{I-vers}
\ee
which should be understood as an expansion in powers of
$G^b_u(M) \Delta$,
\be
\Psi_P({\bf p}) = \Psi_{\bf 0}({\bf p}) 
+\int\frac{d{\bf k}}{(2\pi)^3}\la {\bf p}|G^b_u(M) \Delta|{\bf k}\ra 
\Psi_{\bf 0}({\bf k}) 
+\int\frac{d{\bf k}}{(2\pi)^3} \la {\bf p}|G^b_u(M) \Delta G^b_u(M)
\Delta|{\bf k}\ra\Psi_{\bf 0}({\bf k}) +\cdots, \eqn{psin*}
\ee
where  
\be
G^b_u(E,{\bf p}',{\bf p})=G_u(E,{\bf p}',{\bf p})
-\frac{\Psi_{\bf 0}({\bf p}')\bar\Psi_{\bf 0}({\bf p})}{E-M}.
\eqn{Gb(E)}
\ee
$G_u(E,{\bf p}',{\bf p})$ $[\equiv\la{\bf p}'|G_u(E)|{\bf p}\ra]$
is the Green function corresponding to the homogeneous equation
(\ref{rel-Psi-0}) and satisfying the inhomogeneous equation
\begin{align}
&\left({E}-\sqrt{m_1^2+{\bf p}^{\prime\,2}}
-\sqrt{m_2^2+{\bf p}^{\prime\,2}}\right)
G_u(E,{\bf p}',{\bf p})-\int \frac{d{\bf k}}{(2\pi)^3}
V({E},{\bf 0},{\bf p}',{\bf k})G_u(E,{\bf k},{\bf p})\nn
&\ \ \ \ \ \ \ \ \ \ =(2\pi)^3\delta^3({\bf p}'-{\bf p}). 
\eqn{Gu-E}
\end{align}
The wave function $\Psi_{\bf 0}$ in \eq{Gb(E)} is normalized
in a similar way as in Eq. (\ref{Renorm}), but in the c.m. frame:  
\be
\int \frac{d{\bf p}}{(2\pi)^3}\frac{d{\bf k}}{(2\pi)^3}
\bar\Psi_{\bf 0}({\bf p}) \left.
\left[(2\pi)^3\delta^3({\bf p}-{\bf k})
-\frac{\partial V(E,{\bf 0},{\bf p},{\bf k})}{\partial E}\,\right]
\Psi_{\bf 0}({\bf k})\right|_{E=M}
=1.
\eqn{norm-der}
\ee
The operator $\Delta$ is defined as
\begin{align}
&\la{\bf p}'|\Delta|{\bf p}\ra=V(P,{\bf p}',{\bf p})
-V(M,{\bf 0},{\bf p}',{\bf p})
\nn
&+(2\pi)^3\delta^3({\bf p}'-{\bf p})\left(M-\sqrt{m_1^2+{\bf p}^2}
-\sqrt{m_2^2+{\bf p}^2}-P^0+\sqrt{m_1^2+{\bf p}_1^2}
+\sqrt{m_2^2+{\bf p}_2^2}\right),
\eqn{Delta}
\end{align}
where ${\bf p}_1=\frac{m_1}{m_1+m_2}{\bf P}+{\bf p}$,
${\bf p}_2=\frac{m_2}{m_1+m_2}{\bf P}-{\bf p}$ and
$P^0=\sqrt{M^2+{\bf P}^2}$.
The wave function (\ref{psin*})
is expressed directly in terms of those $NN$ potentials that are
studied in detail and used for the description of the few-body
scattering processes in both approaches, based on the TOPT with
energy-dependent potentials and on the unitary 
transformed Hamiltonian with energy-independent potentials.
\par
It should be emphasized here that the choice of the nonrelativistic
relative momentum provides cancellation of the $v^2$-order terms in
the kinematic part of $\Delta$, \eq{Delta}, as it should be in
the nonrelativistic limit. The $v^4$-order terms are
$\frac{{\bf P}^2{\bf p}^2+ {2({\bf P}.{\bf p})^2}}{M^3}$ (counting also
$\mathbf{p}$ of order $v$) as is shown below in \eq{4th};
it will be shown that the choice of the relativistic relative
momentum provides  cancellation of a part of these $v^4$ terms.
\par
Due to \eq{rel-Psi-0}, along with the normalization condition
(\ref{norm-der}) for the wave function at rest, $\Psi_{\bf 0}$,
\eq{psin*} determines $\Psi_P$ unambiguously; 
it may be more convenient for the latter, which is normalized
as in \eq{norm-der}, to continue to be normalized according
to \eq{Renorm}, so that it could be used directly in the
description of the processes where the bound states are involved
\cite{KBrenorm}.
Indeed, for example, a bound state current in the TOPT is usually
expressed in terms of wave functions normalized by 
\eq{Renorm} as follows:  
\be
\la {\bf P}'|J^\mu |{\bf P}\ra
=\int \frac{d{\bf p}}{(2\pi)^3}\frac{d{\bf k}}{(2\pi)^3}
\bar\Psi_{P'}({\bf p})\Gamma^\mu(P',P,{\bf p},{\bf k})\Psi_P({\bf k}),
\eqn{curr}
\ee
where $\Gamma^\mu(P',P,{\bf p},{\bf k})$ is the five-point amplitude
corresponding to the sum of all $NN$ diagrams which are irreducible
in the in- and out-going channels. 
\par
As explained in \cite{epelbaum}, the approach based on the MUT method
deals with expressions of the currents that are expressed through
the unitary transformed bound state vectors,  
$|^u{\bf P}\ra={\cal U} |{\bf P}\ra$, which have only two-nucleon
Fock-state components, in the form
\be 
\la {\bf P}'|J^\mu |{\bf P}\ra=\la^u {\bf P}'|
{\cal U} J^\mu {\cal U}^{\dagger}|^u{\bf P}\ra.
\ee
The corresponding wave function
$\la{\bf p}_1,{\bf p}_2|^u{\bf P}\ra=
(2\pi)^3\delta^3({{\bf p}_1+{\bf p}_2-\bf P})
\sqrt{\Omega({\bf p}_1^{},{\bf p}_2^{},{\bf P})}\Psi_P^u({\bf p})$
[$\Omega$ defined in \eq{Omg}] is normalized as
\be \lb{e23}
\int d{\bf p}\bar\Psi_P^u({\bf p}) \Psi_P^u({\bf p})=1.
\ee
Notice that this is consistent with \eq{Renorm}, because the
potentials in the MUT do not depend on the on-shell energy $P^0$,
so that $\partial V(P,{\bf p},{\bf k})/\partial P^0=0$ there.
\par

\section{Lorentz boost transformation of the wave function} \lb{s3}  

The choice of \eq{rel-Psi-0} as the unperturbed equation is manifestly
the simplest one, since most of the bound state calculations are done
in the cm frame. However, other choices remain possible. This is the
case when some parts of the higher-order terms of the PT are
known or guessed. Then the latter could also be incorporated into
the unperturbed equation and a perturbation theory could be developed
around the new choice, which might improve or accelerate the
convergence of the PT series.
\par
The latter procedure could be applied 
to the kinematic part of the Lorentz boost operator, which
naturally accompanies the cm wave function transformation.
Based on the one-meson-exchange model \cite{Friar}, an approximate
expression for the Lorentz boost operator has been proposed in the
literature \cite{Schiavilla,UT-boost,moller}, providing the wave
function transformation to the moving reference frame with velocity
$\mathbf{v}=\mathbf{P}/M$:
\be
\Psi_{\bf P}({\bf p})\simeq (1-v^2)^{\frac{1}{4}}
\left(1 -\frac{i{\bf v}\cdot [\,({\bsig}_1-{\bsig}_2)
\times {\bf p}\,]}{4m}\right)
\Psi_{\bf 0}\left(\overrightarrow{\Lambda^{-1} p}\right),
\hspace{1cm}
\overrightarrow{\Lambda^{-1} p}=
{\bf p}-\frac{{\bf v}\cdot{\bf p}}{2}{\bf v}.
\eqn{Schi}
\ee
Throughout the paper, $\Lambda^{-1}$ is the Lorentz transformation
which brings the total four-momentum $P$ to its rest frame:
$\Lambda^{-1}P=(M,{\bf 0})$; $\overrightarrow{\Lambda^{-1} p}$ in
\eq{Schi} is the 3D projection of the $\Lambda^{-1}$ transformation
of a four-dimensional (4D) vector $p$, whose time-component is
$p^0=\frac{{\bf p\cdot P}}{P^0}$.
Formula (\ref{Schi}) is considered for the two-nucleon case, $m$
representing the nucleon mass. 
\par
We shall prove in this section, with general arguments, independent
of any model prescription, that the transformation (\ref{Schi})
indeed represents the kinematic part of the Lorentz boost
transformation of the bound
state wave function to order $v^4$, where the relative momentum
$\mathbf{p}$ is also counted of order $v$ with respect to $m$, as is
the case in nonrelativistic approximations.
\par
To derive the relative momentum transformation,
$\overrightarrow{(\Lambda^{-1} p)}$, featured in \eq{Schi}, it is
sufficient to consider the case of spinless nucleons. In the
approaches based on TOPT, the two-nucleon bound state wave function
$\Psi_{\bf P}({\bf p})$ is strictly related to the field-theoretic
bound state eigenvector $|P\ra\ra$ of the quantized four-momentum
operator with interaction by relation (\ref{e2*}):
\be
(2\pi)^3\delta^3({\bf p}_1+{\bf p}_2-{\bf P})
\sqrt{\Omega({\bf p}_1^{},{\bf p}_2^{},{\bf P})}
\Psi_{\bf P}({\bf p})
=\la p_1,p_2|P\ra\ra=\la p_1,p_2|U_\Lambda |{\bf P}={\bf 0}\ra\ra,
\eqn{37}
\ee
where 
$| p_1,p_2\ra$ is the two-nucleon Fock eigenstate of the
four-momentum operator without interaction and $U_{\Lambda}$ the
unitary operator of Lorentz transformations with account of
interaction.
\par  
Although the discussion in this section is based on TOPT, the
same analysis also applies to the approach based on the MUT with
the two-nucleon bound-state wave function 
$\Psi^u_{\bf P}({\bf p})$, defined as follows:
\be
(2\pi)^3\delta^3({{\bf p}_1+{\bf p}_2-\bf P})
\sqrt{\Omega({\bf p}_1^{},{\bf p}_2^{},{\bf P})}
\Psi_{\bf P}^u({\bf p})
=\la p_1,p_2|\,{\cal U}\,|P\ra\ra
=\la p_1,p_2|\,{\cal U}U_\Lambda {\cal U}^{\dagger} {\cal U}\,|
{\bf P}={\bf 0}\ra\ra,
\eqn{37=MUT}
\ee
where ${\cal U}$ is the unitary transformation which block
diagonalises the Poincar\'e group generators, so that
${\cal U}|P\ra\ra$ has only two-nucleon Fock projection and 
${\cal U}U_\Lambda {\cal U}^{\dagger}$ is block diagonal, i.e.,
the matrix elements corresponding to transitions between two-nucleon
states, with and without mesons, respectively, are zero.
\par  
Definition (\ref{37}) implies the normalization condition
(\ref{Renorm})
and $V(P,{\bf p},{\bf k})$ is given by the sum of all two-nucleon
irreducible time-ordered diagrams in the way displayed in Eqs.
(\ref{Psi-W}) and (\ref{Proj}). 
\par  
To consider the leading order (LO) approximation, we  switch-off
the interaction in the full boost operator $U_\Lambda$, so that it
is replaced in \eq{37} by the one without interaction, $U^0_\Lambda$:
\begin{align} 
\la p_1,p_2|P\ra\ra&=\la p_1,p_2|U_\Lambda |{\bf P}={\bf 0}\ra\ra
\simeq \la p_1,p_2|U^0_\Lambda|{\bf 0}\ra\ra
=(\la\la {\bf 0}| U^{0\dagger}_{\Lambda}|p_1,p_2\ra)^*
\nn
&=(\la\la {\bf 0}| U^{0-1}_{\Lambda }|p_1,p_2\ra)^*
=(\la\la {\bf 0}| U^0_{\Lambda^{-1}}|p_1,p_2\ra)^*
=\la\Lambda^{-1}p_1,\Lambda^{-1}p_2|{\bf 0}\ra\ra,
\eqn{38}
\end{align}
where the superscripts "*" and $"^{\dagger}"$
stand for complex conjugation of functions and for hermitian
conjugation of operators and matrices, respectively.
In such a leading order approximation, \eq{38} establishes
the kinematic relation between the wave function of the moving
bound state, $\Psi_{\bf P}({\bf p})$, and that of the bound state
at rest, $\Psi_{\bf 0}({\bf p})$ (in the scalar-constituent case).
Using definition (\ref{37}) of the bound state wave function
$\Psi_{\bf P}({\bf p})$, and assuming $m_1=m_2=m$, one obtains 
\begin{align} \lb{e21}
&(2\pi)^3\delta^3({\bf p}_1+{\bf p}_2-{\bf P})
\sqrt{\Omega({\bf p}_1^{},{\bf p}_2^{},{\bf P})}
\Psi_{\bf P}({\bf p})
=(\la\la {\bf 0}| U^0_{\Lambda^{-1}} |p_1,p_2\ra)^*\nn
&=\la \Lambda^{-1}p_1,\Lambda^{-1}p_2|{\bf 0}\ra\ra
\nn
&=(2\pi)^3\delta^3(\overrightarrow{\Lambda^{-1} p_1}+
\overrightarrow{\Lambda^{-1} p_2}-{\bf 0})
\sqrt{\Omega(\overrightarrow{\Lambda^{-1} p_1},
\overrightarrow{\Lambda^{-1} p_2},{\bf 0})}
\Psi_{\bf {\bf 0}}(\overrightarrow{\Lambda^{-1} p})
\nn
&\simeq
(2\pi)^3\delta^3({\bf p}_1+{\bf p}_2-{\bf P})
\left(1-\frac{v^2}{2}\right)^{-1}
\sqrt{\Omega(\overrightarrow{\Lambda^{-1} p_1},
\overrightarrow{\Lambda^{-1} p_2},{\bf 0})}
\Psi_{\bf {\bf 0}}(\overrightarrow{\Lambda^{-1} p}),
\end{align}
where the factor $\left(1-\frac{v^2}{2}\right)^{-1}$ comes from the
relation  
$\delta^3(\overrightarrow{\Lambda^{-1} p_1}
+\overrightarrow{\Lambda^{-1} p_2}-{\bf 0})\simeq
\left(1-\frac{v^2}{2}\right)^{-1}
\delta^3 ({\bf p}_1+{\bf p}_2-{\bf P})$, which is valid in the
$v^2$ approximation (cf. Appendix \ref{fact-del}).
In \eq{e21} we have used the general rule of Lorentz transformations,
$U_\Lambda|{\bf P}\ra\ra=
|\overrightarrow{\Lambda P}\ra\ra$ and 
$U^0_\Lambda|{\bf p}\ra=
|\overrightarrow{\Lambda p} \ra$, for the eigenstates 
$|{\bf P}\ra\ra$ and $|{\bf p}\ra$
of the quantized four-momentum operators with and without interaction,
respectively, and Lorentz transformation of 
the nucleon on-mass-shell momenta,
$p_i=\left(\sqrt{m^2+{\bf p}_i^2},{\bf p}_i\right)$, $i=1,2$,
\be
\overrightarrow{(\Lambda^{-1} p_i)}
={\bf p}_i  - p_i^0 {\bf v}
+\frac{{\bf v}\cdot{\bf p}_i }{1+v^0}{\bf v} 
={\bf p}_i  - \left(m+\frac{{\bf p}_i^{\,2}}{2m}\right) {\bf v} +
\frac{{\bf v}\cdot{\bf p}_i }{2} {\bf v} +{\cal O}(v^3), \eqn{38**}
\ee
leading to the $\sqrt{1-v^2}$ contraction of the longitudinal
projection of the relative momentum, 
${\bf p}=\frac{1}{2}({\bf p}_1-{\bf p}_2)$,
\be
\overrightarrow{\Lambda^{-1} p}\simeq {\bf p}
-\frac{{\bf  v}\cdot{\bf  p} }{2}
{\bf v}:\hspace{1cm}
(\overrightarrow{\Lambda^{-1} p})_\perp
={\bf p}_\perp, \hspace{1cm}
(\overrightarrow{\Lambda^{-1} p})_\parallel
\simeq {\bf p}_\parallel\sqrt{1-v^2},  \eqn{34*}
\ee
where $\perp$ and $\parallel$ stand for perpendicular and parallel
projections with respect to ${\bf v}\sim {\bf P}$.
We have also neglected the higher-order terms,
$\frac{{\bf v}\cdot({\bf  p}_1+{\bf p}_2)}{2}{\bf v}
+\frac{({\bf p}_1^{\,2}+{\bf p}_2^{\,2})}{2}{\bf v}+\cdots
={\cal O}(v^3)$, in the $\delta$ function.
\par
On comparison of the left-hand and right-hand sides of Eqs. 
(\ref{e21}), one obtains the relationship between the wave function
of the moving frame and the wave function of the c.m. frame:
\be \lb{e24}
\Psi_{\bf P}({\bf p}) =
\frac{\sqrt{\Omega(\overrightarrow{\Lambda^{-1} p_1},
\overrightarrow{\Lambda^{-1} p_2},{\bf 0})}}
{\sqrt{\Omega({\bf p}_1,{\bf p}_2,{\bf P})}}
\left(1-\frac{v^2}{2}\right)^{-1}
\Psi_{\bf 0}(\overrightarrow{\Lambda^{-1} p})
\simeq (1-v^2)^{1/4} \Psi_{\bf 0}(\overrightarrow{\Lambda^{-1} p}),
\ee
where the ratio of square-root factors depends on ${\bf P}$ and
${\bf p}$, but at order $v^2$ it can be
expressed in terms of only $v^2$ (cf. Appendix \ref{fact-del}):
\be \lb{e29}
\frac{\sqrt{\Omega(\overrightarrow{\Lambda^{-1} p_1},
\overrightarrow{\Lambda^{-1} p_2},{\bf 0})}}
{\sqrt{\Omega({\bf p}_1,{\bf p}_2,{\bf P})}}
\simeq (1-3v^2)^{1/4}.
\ee
The latter factor, multiplied by $\left(1-\frac{v^2}{2}\right)^{-1}$
in \eq{e24}, results into the 
factor $(1-v^2)^{1/4}$, which appears in the right-hand side.
\par
To consider the spin-1/2 constituent case, we use in Eqs.
(\ref{38}) the definition  of the Wigner rotation matrices 
${\cal D}(\Lambda^{-1}p,p)_j^i$ \cite{Weinb63,Gasior66},
\be
U^0_{\Lambda^{-1}}|p,i\ra
=\sum_j{\cal D}(\Lambda^{-1}p,p)_j^i|\Lambda^{-1}p,j\ra, \eqn{P-0-11}
\ee
and we note that the operation $|P\ra\ra=U_\Lambda |{\bf P}={\bf 0}
\ra\ra$ in Eqs. (\ref{38}) does not involve a Wigner rotation matrix,
because 
$[{\cal D}]^i_j(R(\Lambda\lambda_0,\lambda_0))=\delta^i_j$
for any spin when $\vec\lambda_0=0$; although well known, this result
can be checked directly in the spin-$1/2$ case using definition
(\ref{defD}).
Using then the definition of the LO approximation of the bound state
wave function, $\Psi^{D\bf v}_{i_1,i_2}({\bf p})$, 
\begin{align}
&(2\pi)^3\delta^3({\bf p}_1+{\bf p}_2-{\bf P})
\sqrt{\Omega({\bf p}_1^{},{\bf p}_2^{},{\bf P})}
\Psi^{D\bf v}_{i_1,i_2}({\bf p})=
\la p_1i_1,p_2i_2|U^0_\Lambda|{\bf 0}\ra\ra  
\nn
&\ \ \ \ \ \
=(\la\la {\bf 0}| U^0_{\Lambda^{-1} }|p_1i_1,p_2i_2\ra )^* 
\nn
&\ \ \ \ \ \
=\left(\sum_{j_1,j_2}{\cal D}(\Lambda^{-1}p_1,p_1)^{i_1}_{j_1}
{\cal D}(\Lambda^{-1}p_2,p_2)^{i_2}_{j_2}\la\la {\bf 0}
|\Lambda^{-1}p_1j_1,\Lambda^{-1}p_2j_2\ra 
\right)^*
\nn
&\ \ \ \ \ \
=\sum_{j_1,j_2}{\cal D}^*(\Lambda^{-1}p_1,p_1)^{i_1}_{j_1}
{\cal D}^*(\Lambda^{-1}p_2,p_2)^{i_2}_{j_2}
\la\Lambda^{-1}p_1j_1,\Lambda^{-1}p_2j_2|{\bf 0} \ra\ra 
\nn
&\ \ \ \ \ \
=(2\pi)^3\delta^3(\overrightarrow{\Lambda^{-1} p_1}+
\overrightarrow{\Lambda^{-1} p_2}-{\bf 0})
\sqrt{\Omega(\overrightarrow{\Lambda^{-1} p_1},
\overrightarrow{\Lambda^{-1} p_2},{\bf 0})}
\nn
&\ \ \ \ \ \ \ \ \ \ \times \sum_{j_1,j_2}{\cal D}^*
(\Lambda^{-1}p_1,p_1)^{i_1}_{j_1} {\cal D}^*
(\Lambda^{-1}p_2,p_2)^{i_2}_{j_2} 
\Psi^{\bf 0}_{j_1,j_2}(\overrightarrow{\Lambda^{-1} p}),
\eqn{41}
\end{align}
an approximate (kinematic) relation is established between the wave
functions of the moving and resting bound state, 
\begin{align}
\Psi^{D\bf v}_{i_1,i_2}({\bf p})&=(1-v^2)^{1/4}
\sum_{j_1,j_2}{\cal D}^{\dagger}
(\Lambda^{-1}p_1,p_1)_{i_1}^{j_1} {\cal D}^{\dagger}
(\Lambda^{-1}p_2,p_2)^{j_2}_{i_2}
\Psi^{\bf 0}_{j_1,j_2}(\overrightarrow{\Lambda^{-1} p}),  
\nn
\Psi_{D\bf v}^{\dagger i_1,i_2}({\bf p})
&=(1-v^2)^{1/4}\sum_{j_1,j_2}\Psi_{\bf 0}^{\dagger j_1,j_2}
(\overrightarrow{\Lambda^{-1} p})
{\cal D}(\Lambda^{-1}p_1,p_1)_{j_1}^{i_1}
{\cal D}(\Lambda^{-1}p_2,p_2)_{j_2}^{i_2}.
\eqn{+form}
\end{align}
The origin of the factor $(1-v^2)^{1/4}$ is explained in
Eqs. (\ref{e24}) and (\ref{e29}); the lower and upper spin indices
conveniently distinguish
between "covariant" and "contravariant" wave functions; we also
have the properties
${\cal D}^{\dagger}(\Lambda^{-1}p_1,p_1)^{i_1}_{j_1}={\cal D}^*
(\Lambda^{-1}p_1,p_1)_{i_1}^{j_1}$.
Equations (\ref{+form}) can be written in symbolic form as
\begin{align}
\Psi^{D\bf v}({\bf p})
&={\cal D}^{\dagger}(\Lambda^{-1}p_1,p_1){\cal D}^{\dagger}
(\Lambda^{-1}p_2,p_2)\Psi_{\bf 0}(\overrightarrow{\Lambda^{-1} p}), 
\nn 
\Psi_{D\bf v}^{\dagger}({\bf p})
&=\Psi_{\bf 0}^{\dagger}(\overrightarrow{\Lambda^{-1} p})
{\cal D}(\Lambda^{-1}p_1,p_1){\cal D}(\Lambda^{-1}p_2,p_2).
\eqn{sym}
\end{align}
${\cal D}^{\dagger}(\Lambda^{-1}p_1,p_1)_{j_1}^{i_1}
{\cal D}^{\dagger}(\Lambda^{-1}p_2,p_2)_{j_2}^{i_2}$ is calculated
in Appendix \ref{Wigner} [Eq. (\ref{tr=for-inv})]. The relationship
between the two wave functions takes the form  
\be
\Psi^{D\bf v}({\bf p})=(1-v^2)^{1/4}\,\left(1-\frac{i{\bf v}\cdot
[\,({\bsig}_1-{\bsig}_2)\times {\bf p}\,]}{4m}\right)
\Psi_{\bf 0}(\overrightarrow{\Lambda^{-1} p}),
\eqn{final*}
\ee
where  $\overrightarrow{\Lambda^{-1} p}
\simeq{\bf p}-\frac{{\bf v}\cdot{\bf p}}{2}{\bf v}$,
as in \eq{34*}.
The wave function (\ref{final*}) thus reproduces \eq{Schi}, used as
the approximate boost in the literature
\cite{UT-boost,moller,Schiavilla}.
\par
A consistency check can be considered, in the spinless case and
for energy-independent potentials, by
comparing the normalization conditions of the two wave functions,
as given by Eqs. (\ref{Renorm}) [or \eq{e23}] and (\ref{norm-der}).
Replacing in (\ref{Renorm}) $\Psi_P({\bf p})$ by its
expression given in Eq. (\ref{final*}) (without the $\sigma$
matrices), making the change of variable from ${\bf p}$ to
$\overrightarrow{\Lambda^{-1} p}$ and using
$d{\bf p}=(1-v^2)^{-1/2}\,d(\overrightarrow{\Lambda^{-1} p})$, one finds
back Eq. (\ref{norm-der}). This shows that for the perturbation theory
where the lowest-order wave function is (\ref{final*}) (or its spinless
analog) the normalization condition will be developed along \eq{Renorm}.
\par
Although \eq{final*} and what follows in Sec. \ref{s4} 
consider the case of equal masses for the nucleons, there
is no difficulty to deal  with the cases of different masses and
few-particle systems.
For example, in the case of a three-particle system with different
masses, the boost transformation matrix for the bound state wave
function of the three particles would have a form generalizing
\eq{tr=for-inv}: 
\begin{align}
&{\cal D}^{\dagger}(\Lambda^{-1}p_1,p_1)
{\cal D}^{\dagger}(\Lambda^{-1}p_2,p_2)
{\cal D}^\dagger(\Lambda^{-1}p_3,p_3)
=1-i\vv\cdot\left(\frac{[\bsig_1\times \pp_1]}{4m_1}
+\frac{[\bsig_2\times \pp_2]}{4m_2}
+\frac{[\bsig_2\times \pp_3]}{4m_3} \right), \nn 
&
\eqn{tr=for-inv*}
\end{align}
where two relative momenta could be chosen using the relation
$\pp_1+\pp_2+\pp_3={\bf P}$.
\par

\section{ Generalized form of the unperturbed equation} \lb{s4}

One can now use $\Psi^{D\bf v}({\bf p})$ [\eq{final*}] as a LO
approximation for \eq{cor-PT}, instead of $\Psi_{\bf 0}({\bf p})$.
This leads to a rearrangement of the series (\ref{psin*}).
The equation satisfied by $\Psi^{D\bf v}({\bf p})$ follows from
\eq{rel-Psi-0} (cf. Appendix \ref{appb} for the details of the
derivation):
\begin{align}
\left(M-2\sqrt{m^2+(\overrightarrow{\Lambda^{-1} p})^2}\right)
&\Psi^{D\bf v}({\bf p}) \nn
&=(1-v^2)^{1/2}\int \frac{d{\bf k}}{(2\pi)^3}D({\bf v}, {\bf p})
V_0(\overrightarrow{\Lambda^{-1} p},\overrightarrow{\Lambda^{-1} k})
D^{\dagger}({\bf v}, {\bf k})
\Psi^{D\bf v}({\bf k}),  \eqn{PsiM}
\end{align}
where  $D({\bf v},{\bf p})=
1-\frac{i{\bf v}\cdot
[\,({\bsig}_1-{\bsig}_2)\times {\bf p}\,]}{4m}$
and $V_0({\bf p},{\bf k})=V(M,{\bf 0},{\bf p},{\bf k})$.
Then the rearranged PT takes the form
\be
\Psi_P({\bf p})=\int \frac{d{\bf k}}{(2\pi)^3}
\la {\bf p}|[1-G^{b{\bf v}}_u(M) \Delta]^{-1}|
{\bf k}\ra \Psi^{D\bf v}({\bf k}),  \eqn{11222}
\ee
where 
\be
G^{b\bf v}_u(E,{\bf p},{\bf k})=G^{\bf v}_u(E,{\bf p},{\bf k})
-\frac{\Psi^{D\bf v}({\bf p})\bar\Psi^{D\bf v}({\bf k})}{E-M}
=(1-v^2)^{1/2}
D({\bf v},{\bf p})G^b_u(E,{\bf p},{\bf k})D^{\dagger}({\bf v}, {\bf k}).
\ee
$G^{\bf v}_u(E, {\bf p},{\bf k})$
$[\equiv (1-v^2)^{1/2}D({\bf v}, {\bf p})G_u(E,{\bf p},{\bf k})
D^{\dagger}({\bf v},{\bf k})]$ is the Green function corresponding to  
\eq{PsiM},
\be
\left(M-2\sqrt{m^2+(\overrightarrow{\Lambda^{-1} p})^2}\right)
G^{\bf v}_u(E,{\bf p},{\bf p}')
-\int \frac{d{\bf k}}{(2\pi)^3}V^{\bf v}_0({\bf p},{\bf k})
G^{\bf v}_u(E,{\bf k},{\bf p}')=(2\pi)^3\delta^3({\bf p}-{\bf p}'),
\eqn{56}
\ee
where  $V^{\bf v}_0({\bf p},{\bf k})=(1-v^2)^{1/2}
D({\bf v}, {\bf p})
V_0(\overrightarrow{\Lambda^{-1} p},\overrightarrow{\Lambda^{-1} k})
D^{\dagger}({\bf v}, {\bf k})$ and
\begin{align}
 & \la{\bf p}'|\Delta|{\bf p}\ra=
 V(P,{\bf p}',{\bf p})-V^{\bf v}_0({\bf p}',{\bf p})
\nn
&+ (2\pi)^3\delta^3({\bf p}'-{\bf p})
\left(M-2\sqrt{m^2+(\overrightarrow{\Lambda^{-1} p})^2}-P^0
+\sqrt{m^2+{\bf p}_1^2}+\sqrt{m^2+{\bf p}_2^2}\right).
\eqn{Delta-n}
\end{align}
One notices that the choice of the new LO approximation wave
function (\ref{final*}) reduces the interaction part of the
perturbation (\ref{Delta-n}) to\\
$V(P,{\bf p}',{\bf p})-V^{\bf v}_0({\bf p}',{\bf p})
=V(P,{\bf p}',{\bf p})-(1-v^2)^{1/2} D({\bf v}, {\bf p}')
V_0(\overrightarrow{\Lambda^{-1} p'},\overrightarrow{\Lambda^{-1} p})
D^{\dagger}({\bf v}, {\bf p})$. 
The kinematic part of the perturbation (\ref{Delta-n}) yields
\be
\left(P^0-\sqrt{m^2+{\bf p}_1^2}-\sqrt{m^2+{\bf p}_2^2}\right)
-\left(M-2\sqrt{m^2+(\overrightarrow{\Lambda^{-1} p})^2}\right)
=\frac{v^2{\bf p}^2}{M}+{\cal O}(v^6),  \eqn{scal-red}
\ee
where we used
$(\overrightarrow{\Lambda^{-1} p})^2={\bf p}_\perp^2
+(1-v^2){\bf p}_\parallel^2={\bf p}^2-({\bf p}{\bf v})^2$;
this is smaller than that of (\ref{Delta}), as expected,
\be
\left(P^0-\sqrt{m^2+{\bf p}_1^2}-\sqrt{m^2+{\bf p}_2^2}\right)
-\left(M-2\sqrt{m^2+{\bf p}^2}\right)
=\frac{v^2{\bf p}^2+ 2({\bf v}\cdot{\bf p})^2}{M}+{\cal O}(v^6).
\eqn{4th}
\ee
Boosting the relative momentum in the LO approximation wave function,
$\Psi_0(\overrightarrow{\Lambda^{-1} p})$, has resulted in the elimination
of a part of the $v^4$-order contribution, $2({\bf v}{\bf p})^2$, to
\eq{4th}. As the approximate transformation (\ref{final*}) has been
used in the literature, \cite{UT-boost,moller,Schiavilla}, it is
worth mentioning that, for consistency, the remaining $v^4$ corrections
should also be taken into account; our present approach, which considers
the corrective effects in a systematic way, incorporates all of them.
\par
In  covariant approaches, such differences can of course be
eliminated completely. 
\par
The PT (\ref{11222}) may be technically more convenient if written
in a different form upon the use of 
$\Psi^{D\bf v}({\bf p})=(1-v^2)^{1/4}
D({\bf v},{\bf p})\Psi_{\bf 0}(\overrightarrow{\Lambda^{-1} p})$: 
\begin{align}
\Psi_P({\bf p})=&\int \frac{d{\bf k}}{(2\pi)^3}
\la {\bf p}|[1-G^{b{\bf v}}_u(M)
\Delta]^{-1}|{\bf k}\ra \Psi^{D\bf v}({\bf k})
\nn
=&(1-v^2)^{1/4}\int \frac{d{\bf k}}{(2\pi)^3}
\la{\bf p}|[1-DG^{b}_u(M)D^{\dagger}
\Delta]^{-1}|{\bf k}\ra
D({\bf v},{\bf k})\Psi_{\bf 0}(\overrightarrow{\Lambda^{-1} k})
\nn
=&(1-v^2)^{1/4}D({\bf v}, {\bf p})
\int \frac{d{\bf k}}{(2\pi)^3}\la {\bf p}|[1-G^{b}_u(M)D^{\dagger}
\Delta D]^{-1}|{\bf k}\ra 
\Psi_{\bf 0}(\overrightarrow{\Lambda^{-1} k}),
\end{align}
where 
\begin{align}
D^{\dagger} \Delta D&=D^{\dagger}({\bf v},{\bf p})V(P,{\bf p},{\bf k})
D({\bf v}, {\bf k}) -(1-v^2)^{1/2} 
V_0(\overrightarrow{\Lambda^{-1} p},\overrightarrow{\Lambda^{-1} k})
\nn
&+ (2\pi)^3\delta^3({\bf p}-{\bf k})
\left(M-2\sqrt{m^2+(\overrightarrow{\Lambda^{-1} p})^2}
-P^0+\sqrt{m^2+{\bf p}_1^2}+\sqrt{m^2+{\bf p}_2^2}\right).
\end{align} 
\par
While $\Psi^{D\bf v}({\bf p})$ is the LO wave function of the
rearranged PT (\ref{11222}), it can be reproduced in the NLO of the
initial version (\ref{I-vers}). Indeed, as it is the solution of
\eq{PsiM}, its potential
$V^{\bf v}_0({\bf p},{\bf k})=(1-v^2)^{1/2} D({\bf v},{\bf p})
V_0(\overrightarrow{\Lambda^{-1} p},\overrightarrow{\Lambda^{-1} k})
D^{\dagger}({\bf v},{\bf k})$ should be a part of the full potential
featured in the PT (\ref{I-vers}). The latter, written in
(\ref{Delta}) with the approximations $V(P,{\bf p}',{\bf p})
\simeq V^{\bf v}_0({\bf p},{\bf k})$,
$P^0-\sqrt{m^2+{\bf p}_1^2}-\sqrt{m^2+{\bf p}_2^2}
= M-2\sqrt{m^2+(\overrightarrow{\Lambda^{-1} p})^2} +{\cal O}(v^4)$, i.e.,
\begin{align}
&\la{\bf p}'|\Delta|{\bf p}\ra\simeq
\la{\bf p}'|\Delta^{D\bf v}|{\bf p}\ra 
\nn
&=V^{\bf v}_0({\bf p}',{\bf p})
-V(M,{\bf 0},{\bf p}',{\bf p})
+(2\pi)^3\delta^3({\bf p}'-{\bf p})\left( -2\sqrt{m^2+{\bf p}^2}
+2\sqrt{m^2+(\overrightarrow{\Lambda^{-1} p})^2}\right),
\eqn{DeltaDv}
\end{align}
 gives
\be
\Psi_P({\bf p}) \simeq \int \frac{d{\bf k}}{(2\pi)^3}
\la {\bf p}|[1-G^b_u(M)
\Delta^{D\bf v}]^{-1}|{\bf k}\ra \Psi_{\bf 0}({\bf k}).
\eqn{rhs} 
\ee
One verifies that the right-hand-side of \eq{rhs} is a solution
of \eq{PsiM} and thus is $\Psi^{D\bf v}({\bf p})$. 
\par
 
\section{Conclusion} \lb{s5}  
 
We have presented the perturbation theory for the evaluation of the
corrections due to the motion of the bound states involved in
scattering processes. This completes the developments with systematic
approaches to the chiral EFT for nucleons, undertaken for more than
three decades with high precision studies.
\par 
The practical convenience of our PT, expressed in the form of
\eq{psin*}, or of its variants of Sec. \ref{s4}, lies in the fact
that it is formulated in terms of the potentials that are
determined and used for the analysis of the scattering amplitudes
in both TOPT and MUT approaches.
\par
The present study could also be extended without difficulty to the
case of unequal masses and to the case of few-nucleon systems.
\par

\begin{acknowledgments}
A.N.K. and H.S. thank the Erasmus+ MIC Program of Paris-Saclay
University for financial support. They also are grateful to
Sergey Barsuk for his help in finalizing that project.
A.N.K. was partially supported by the Shota Rustaveli National Science
Foundation (Grant No. FR-23-856).
\end{acknowledgments}
 
\begin{appendix}

\section{Emergence of ${\bf v^2}$-dependent factors}
\label{fact-del}

We integrate the first and fourth expressions of Eqs. (\ref{e21})
with respect to the variable ${\bf X}={\bf p}_1+{\bf p}_2$; note that
in the calculations below the total momentum ${\bf P}$ 
and the relative momentum ${\bf p}=({\bf p}_1-{\bf p}_2)/2$
of the state are considered as fixed.
We have
\begin{align} \lb{ae1}
&\sqrt{\Omega({\bf p}_1^{},{\bf p}_2^{},{\bf P})}
\Psi_{\bf P}({\bf p})
=\int \frac{d{\bf X}}{(2\pi)^3}\la p_1,p_2|P\ra\ra
\nn
&\simeq \int \frac{d{\bf X}}{(2\pi)^3}
(2\pi^3)\delta^3
(\overrightarrow{\Lambda^{-1} p_1}+\overrightarrow{\Lambda^{-1} p_2}
-{\bf 0})
\sqrt{\Omega(\overrightarrow{\Lambda^{-1} p_1},
\overrightarrow{\Lambda^{-1} p_2},{\bf 0})}
\Psi_{\bf 0}(\overrightarrow{\Lambda^{-1} p}).
\end{align}
On the other hand,
\begin{align}
&\delta^3
(\overrightarrow{\Lambda^{-1} p_1}+\overrightarrow{\Lambda^{-1} p_2}
-{\bf 0})
\simeq\delta^3
\Big({\bf p}_1-(m+\frac{{\bf p}_1^2}{2m}){\bf v}+
\frac{{\bf v}\cdot{\bf p}_1}{2}{\bf v}
+{\bf p}_2-(m+\frac{{\bf p}_2^2}{2m}){\bf v}+
\frac{{\bf v}\cdot{\bf p}_2}{2}{\bf v}\Big)
\nn
&=\delta^3
\Big({\bf X}-(m+\frac{{\bf p}_1^2}{2m}){\bf v}+
\frac{{\bf v}\cdot{\bf X}}{2}{\bf v}
-(m+\frac{{\bf p}_2^2}{2m}){\bf v}\Big)
\nn
&=\delta^3
\Big({\bf X}+\frac{{\bf v}\cdot{\bf X}}{2}{\bf v}-2m{\bf v}-
\frac{{\bf X}^2+4{\bf p}^2}{4m}{\bf v}\Big)
=\delta^2({\bf X}_\perp)
\delta
\Big(X_\parallel+\frac{v^2X_\parallel}{2}-2mv-\frac{X_\parallel^2
 +4{\bf p}^2}{4m}v\Big)
\nn
&=\delta^2({\bf X}_\perp)\delta (X_\parallel-\bar X_\parallel)
\Big[\frac{\partial}{\partial X_\parallel}
\Big(X_\parallel+\frac{v^2X_\parallel}{2}-2mv
-\frac{X_\parallel^2+4{\bf p}^2}{4m}v\Big)\Big]^{-1}
\nn
&
=\delta^2({\bf X}_\perp)
\delta (X_\parallel-\bar X_\parallel)
\Big(1+\frac{v^2}{2}-\frac{2X_\parallel}{4m}v\Big)^{-1}
\simeq \delta^2({\bf X}_\perp)
\delta (X_\parallel-\bar X_\parallel)
\Big(1+\frac{v^2}{2}-v^2\Big)^{-1}
\nn
&
=\delta^2({\bf X}_\perp)
\delta (X_\parallel-\bar X_\parallel)
\Big(1-\frac{v^2}{2}\Big)^{-1}.
\end{align}
We used $\bar X_\parallel=Mv+\delta=Mv+\frac{{\bf p}^2}{m}v$
for the solution of equation 
$X_\parallel+\frac{v^2X_\parallel}{2}-2mv
-\frac{X_\parallel^2+4{\bf p}^2}{4m}v=0$:
\be
Mv+\delta+\frac{v^2(Mv+\delta)}{2}-2mv
-\frac{M^2v^2+2mv\delta+4{\bf p}^2}{4m}v=
\delta-\frac{{\bf p}^2}{m}v=0.
\ee
Therefore, the integration with respect to ${\bf X}$ in \eq{ae1}
provides the multiplicative factor\\ $\big(1-\frac{v^2}{2}\big)^{-1}$.
\par
 Derivation of \eq{e29}: 
\begin{align}
&\frac{\sqrt{\Omega(\overrightarrow{\Lambda^{-1} p_1},
\overrightarrow{\Lambda^{-1} p_2},{\bf 0})}}
{\sqrt{\Omega({\bf p}_1,{\bf p}_2,{\bf P})}}
=
\left(
\frac{(m^2+({\bf p}_1-m{\bf v})^2)(m^2+({\bf p}_2-m{\bf v})^2)M^2}
{(m^2+{\bf p}_1^2)(m^2+{\bf p}_2^2)(M^2+{\bf P}^2)}
\right)^\frac{1}{4}
\nn
&=
\left(
\frac{(m^2+({\bf p}_1-m{\bf v})^2)(m^2+({\bf p}_2-m{\bf v})^2)}
{\omega_1^2\omega_2^2(1+v^2)}
\right)^\frac{1}{4}
\nn
&=
\left(
\frac{(\omega_1^2+m^2v^2-2m{\bf p}_1\cdot {\bf v})
(\omega_2^2+m^2v^2-2m{\bf p}_2\cdot {\bf v})}
{\omega_1^2\omega_2^2(1+v^2)}
\right)^\frac{1}{4}
\nn
&\simeq
\left(
\frac{\omega_1^2\omega_2+\omega_2^2(m^2v^2-2m{\bf p}_1\cdot {\bf v})
+\omega_1^2(m^2v^2-2m{\bf p}_2\cdot {\bf v})}
{\omega_1^2\omega_2^2(1+v^2)}\right)^\frac{1}{4}
\nn
&\simeq
\left(
\frac{1+\omega_1^{-2}(m^2v^2-2m{\bf p}_1\cdot {\bf v})
+\omega_2^{-2}(m^2v^2-2m{\bf p}_2\cdot {\bf v})}
{1+v^2}\right)^\frac{1}{4}
\nn
&
\simeq
\left(
\frac{1+m^{-2}(m^2v^2-2m{\bf p}_1\cdot {\bf v}+m^2v^2
-2m{\bf p}_2\cdot {\bf v})}{1+v^2}
\right)^\frac{1}{4}
=\left(
\frac{1+m^{-2}(2m^2v^2-2m{\bf P}\cdot {\bf v})}{1+v^2}\right)^\frac{1}{4}
\nn
&=\left(\frac{1-2v^2}{1+v^2}\right)^\frac{1}{4}
\simeq (1-3v^2)^\frac{1}{4}.
\end{align}

\section{Wigner rotation matrices} \label{Wigner}
 
For the spin-$1/2$ case, the Wigner rotation matrix is defined
as \cite{Weinb63,Gasior66}
\be
[{\cal D}^\frac{1}{2}]^i_j(R(\Lambda p,p))=\bar u_j(\Lambda p)
S(\Lambda)u^i(p), \hspace{1cm} p^2=m^2, 
\eqn{defD}
\ee
where $S(\Lambda)$ is the $4\times 4$ matrix representation of the
Lorenz transformation $\Lambda$ in the Dirac spinor space and
$u^i(p)$, $\bar u_j(p)$ ($i,j=1,2$) are one of the two independent
Dirac spinor solutions and their adjoints, respectively,
\be
S^{-1}(\Lambda) \gamma^\mu S(\Lambda)=\Lambda^\mu_\nu \gamma^\nu,
\hspace{5mm} 
u^i({p}) = \sqrt{\frac{E_p+m}{2m}}\left(\begin{array}{c} 1 \\[2mm]
\displaystyle\frac{\bsig\cdot\bf p}{E_p+m}\end{array}\right)
\chi^i, \hspace{5mm}  
\bar u_j(p) u^i(p)= \delta^i_j. 
\eqn{7**}
\ee
($\chi^i$ is one of the two momentum-independent two-component
independent spinors.)
\par
For the boost transformation, $\Lambda^{-1}v=(1,\vec 0)$, where $v=P/M$,
we have \cite{Gasior66}
\be
S(\Lambda^{-1})=\cosh(\frac{\phi}{2}) -
\left( \begin{array}{cc} 0 &\bsig\cdot\bf n \\
\bsig\cdot\bf n &0 \end{array} \right)\sinh(\frac{\phi}{2}),
\hspace{5mm} {\bf n}=\frac{{\bf v}}{v}, \hspace{5mm}
\sinh\phi=\frac{v}{\sqrt{1-v^2}}.
\ee
To compare it with approximate expressions used in the literature
\cite{UT-boost,moller,Schiavilla}, we consider the $v^2$ approximation,
where $\phi\simeq v$,
\be
S(\Lambda^{-1})=
1+\frac{v^2}{8}
-\frac{1}{2}\left(\begin{array}{cc} 0 &\bsig\cdot\bf v \\
\bsig\cdot\bf v &0 \end{array} \right)
+{\cal O}(v^3)=
\left(1+\frac{v^2}{8}\right)\left[1 - \frac{1}{2}
\left(\begin{array}{cc} 0 &\bsig\cdot\bf v \\
\bsig\cdot\bf v &0 \end{array} \right)\right]
+{\cal O}(v^3).
\ee
We then obtain
\begin{align}
&[{\cal D}^\frac{1}{2}]^i_j(R(\Lambda^{-1} p,p))= 
\bar u_j(\Lambda^{-1} p) S(\Lambda^{-1})u^i(p)
\nn
&=\frac{1}{2m}\sqrt{E_{\Lambda^{-1}p}+m} \sqrt{E_p+m}
\left(1-\frac{{\bf v}^2}{8}-\frac{{\bf p}^2}{4m^2}
+\frac{{\bf v}\cdot{\bf p} +i{\bf v}\cdot[\,{\bsig}
\times{\bf  p}\,]}{4m}\right)_j^i
\nn
&=\left(1+\frac{i{\bf v}\cdot[\,{\bsig}\times{\bf p}\,]}{4m}
\right)_j^i,  \eqn{3345}
\end{align}
where we used $\overrightarrow{\Lambda^{-1} p}={\bf p}-m{\bf v}$,
and
\begin{align}
&\frac{1}{2m}\sqrt{E_{\Lambda^{-1}p}+m} \sqrt{E_p+m}
=\sqrt{1+\frac{({\bf p}-m{\bf v})^2}{4m^2}}
\sqrt{1+\frac{{\bf p}^2}{4m^2}}
\simeq 1+\frac{({\bf p}-m{\bf v})^2+{\bf p}^2}{8m^2},   
\nn
& \left(1+\frac{({\bf p}-m{\bf v})^2+{\bf p}^2}{8m^2}\right)
\left(1 -\frac{{\bf v}^2}{8}   -\frac{{\bf p}^2}{4m^2}
+\frac{{\bf p}\cdot{\bf v}}{4m}\right) \simeq 1.
\end{align}
\par
The boost transformation matrix for the two-nucleon bound state
wave function then takes the form
\begin{align}
&{\cal D}^{\dagger}(\Lambda^{-1}p_1,p_1)
{\cal D}^{\dagger}(\Lambda^{-1}p_2,p_2)
=\left(1-\frac{i{\bf v} \cdot[\,{\bsig}_1\times{\bf p}_1]}{4m}\right)
\left(1-\frac{i{\bf v}\cdot[{\bsig}_2\times {\bf p}_2]}{4m}\right)
\nn
&\ \ \ \ \ \simeq 1-\frac{i{\bf v}\cdot[\,({\bsig}_1-{\bsig}_2)
\times{\bf p}\,]}{4m}.
\eqn{tr=for-inv}
\end{align}
\par

\section{The LO wave equation for moving bound states} \lb{appb}

Using the relationship between the LO wave functions of the moving
bound state and of the one at rest,
$\Psi^{D\bf v}({\bf p})=(1-v^2)^{1/4}D({\bf v},{\bf p})
\Psi_0(\overrightarrow{\Lambda^{-1} p})$,
$D({\bf v}, {\bf p})=1-\frac{i\vec v\cdot[(\vec\sigma_1
-\vec\sigma_2)\times \vec p]}{4m}$ [(\eq{final*})], and the equation
satisfied by $\Psi_{\bf 0}({\bf k})$ [(\eq{rel-Psi-0})], one obtains 
\begin{align}
&\left(M-2\sqrt{m^2+(\overrightarrow{\Lambda^{-1} p})^2}\right)
\Psi^{D\bf v}({\bf p})
=(1-v^2)^{1/4}
D({\bf v}, {\bf p})\left(M-2\sqrt{m^2+(\overrightarrow{\Lambda^{-1} p})^2}
\right)\Psi_0(\overrightarrow{\Lambda^{-1} p})
\nn
&\ \ \ \ \ \ \ \ =(1-v^2)^{1/4}
D({\bf v}, {\bf p})\int 
\frac{d(\overrightarrow{\Lambda^{-1} k})}{(2\pi)^3}
V_0(\overrightarrow{\Lambda^{-1} p},\overrightarrow{\Lambda^{-1} k})
\Psi_0(\overrightarrow{\Lambda^{-1} k})
\nn
&\ \ \ \ \ \ \ \ =(1-v^2)^{1/2}\int \frac{d{\bf k}}{(2\pi)^3}
D({\bf v}, {\bf p})
V_0(\overrightarrow{\Lambda^{-1} p},\overrightarrow{\Lambda^{-1} k})
D^{\dagger}({\bf v}, {\bf k})
\Psi^{D\bf v}({\bf k}),
& \eqn{PsiM*}
\end{align}
where the relation
$d(\overrightarrow{\Lambda^{-1} k})=\sqrt{1-v^2} d{\bf k}$ has been used.
\par

\end{appendix}

\end{document}